\documentclass[aps,prl,superscriptaddress,unsortedaddress,twocolumn,showpacs]{revtex4}

\usepackage{amsfonts}

\usepackage[pdftex]{graphicx}
\usepackage{epstopdf}
\usepackage{amsmath}
\usepackage{amssymb}
\usepackage{amsthm}
\usepackage{tikz}

\definecolor{linkcolor}{rgb}{0,0,0.6}
\usepackage[pdftex,colorlinks=true, pdfstartview=FitV, linkcolor= linkcolor, citecolor= linkcolor, urlcolor= linkcolor, hyperindex=true,hyperfigures=true]{hyperref}

\begin{document}

\title{Energy Transfer between Colloids via Critical Interactions.}
\author{Ignacio A. Mart\'inez }
\email{iamartinez@ucm.es}
\affiliation{Laboratoire de Physique, \'Ecole Normale Sup\'erieure, CNRS UMR5672 46 All\'ee d'Italie, 69364 Lyon, France.}
\affiliation{Departamento de F\' isica At\'omica, Molecular y Nuclear, Universidad Complutense de Madrid, 28040 Madrid, Spain}
\author{Clemence Devailly }
\affiliation{Laboratoire de Physique, \'Ecole Normale Sup\'erieure, CNRS UMR5672 46 All\'ee d'Italie, 69364 Lyon, France.}
\affiliation{SUPA and School of Physics and Astronomy, The University of Edinburgh, JCMB, Peter Guthrie Tait Road, Edinburgh EH9 3FD, UK.}
\author{Artyom Petrosyan}
\affiliation{Laboratoire de Physique, \'Ecole Normale Sup\'erieure, CNRS UMR5672 46 All\'ee d'Italie, 69364 Lyon, France.}
\author{Sergio Ciliberto}
\email{sergio.ciliberto@ens-lyon.fr}
\affiliation{Laboratoire de Physique, \'Ecole Normale Sup\'erieure, CNRS UMR5672 46 All\'ee d'Italie, 69364 Lyon, France.}

\begin{abstract}
{We report the observation of a temperature-controlled synchronization of two Brownian-particles in a binary mixture  close to the critical point of the demixing transition. The two beads are trapped by two optical tweezers whose distance is periodically modulated.      We notice that the motion synchronization  of the two beads appears when the critical temperature is approached. In contrast, when the fluid is far  from its critical temperature, the  displacements of the two beads  are uncorrelated.  Small changes in temperature can radically change the global dynamics of the system.   We show that the synchronisation is induced by the critical Casimir forces. Finally, we present the  measure of  the energy transfers inside the system produced by the critical interaction.  }
\end{abstract}




\maketitle

\section{Introduction}

Synchronization is the coordination of events which allow a system to operate in unison \cite{vicsek2012}. This effect appears in Nature  at any  size, from the smallest systems, as in the quantum world \cite{ozyuzer2007}, to the rotation of binary stars \cite{lamb1983,giuricin1984}, going through something as mundane as the passage of people on public transportation during rush hour. In the specific case of mesoscales, from nanometers to microns, the use of colloidal particles is widely accepted to study synchronization. Different mechanisms have  been proposed to induce  interactions between particles. For example, hydrodynamics \cite{kotar2010,bartlett2001}, magnetic \cite{gao2010} or electric \cite{dobnikar2013}, even Janus colloidal particles have been used to test what the  synchronization may be like at the origin of  structure formations \cite{yan2012}. 
In this article we present a new form of synchronization between two colloidal particles inside a binary mixture close to its mixing critical point. We show that this  synchronization is induced  by the critical Casimir force and  it is temperature-controlled. We also study the energy transfer properties  of this two particles  system using the  stochastic thermodynamics framework
 \cite{seifert2011,seki2010,ciliannrev}.
A binary mixture at the critical concentration is a specific case of a system presenting a second order phase transition. When the mixture approaches the critical temperature, $T_c$, the correlation length  
$\xi$ of its thermal fluctuations diverges, i.e., $\xi=\xi_0 \varepsilon ^{-\nu}$ , where 
$\varepsilon=(T_c-T)/T_c$ is the reduced temperature and $\xi_0$ is the characteristic correlation length of the mixture. The exponent $\nu$ $=$ 0.63 is the universal exponent associated with the transition. If the fluctuating field $\phi$ is confined on length scales comparable to $\xi$  the critical Casimir force between the confining walls appears \cite{fisher1978}. This force presents a great interest in the nanotechnology area due to its long-range nature and to the possibility of being switched on-off in a simple way, because  small changes in the temperature of the system will power those forces.  In the last decade, the number of  experimental studies of such a kind of systems has  increased exponentially, because the possibility of  exploring those fluctuating fields locally, of  managing nanometric systems and of detecting  femtoNewton forces has allowed  the direct observation of these forces a few years ago by Hertlein et al. \cite{hertlein2008}. Indeed the application of critical Casimir force to  nanotechnologies has  been proposed since its theoretical development in the seventies. For example, very recently, aggregation of particles was induced by      {critical Casimir interactions} \cite{faber2013}. The perspectives  of implementing these forces in more complex mechanisms are giant. For example, an  important application could be   the miniaturization of thermodynamic engines to single-molecule devices, which require the development of switches with the ability of being turn on/off in a controllable way.

In our experiment, a dual optical  trap  holds two colloidal particles  inside a binary mixture. The equilibrium positions of the two optical traps are independent, one is kept fixed while the other is periodically moved in order to change the distance between the two particles. The moving trap injects energy into the system and we study the transfer of energy from the \textit{moving particle} to the \textit{fixed one}. The novelty of our experiment is that the particles are able to interact by means of   the {Critical Casimir force} which can be easily activated by  tuning the  temperature of the mixture  close to  $T_{\rm C}$.

\section{Results}

The experiment is carried out in the following way:  first, the two microsized beads $1$ (fixed) and 2 (moving) are optically trapped in the low critical temperature micelles-solvent solution at a stabilized temperature ($\pm 10$ mK) and at a constant trap stiffness, $\kappa_1=\kappa_2=(0.5\pm0.1)$ ${\rm pN/}\mu {\rm m}$. Then, the protocol in the positions of the traps $\Gamma$ is imposed by fixing  the position of the first trap at  $x^T_1=0.000$ $\mu$m while the position $x^T_2(t)$ of the second trap is periodically moved according to the following procedure which lasts $4\tau=2$ s. The trap $2$  is kept for $0\le t \le \tau$ at $x^T_{2}=5.55$ $\mu$m. It is then driven at a  constant speed ($v^T_2=v\simeq$ 1.7 $\mu$m/s) from this static position  to a new  position at $x^T_{2}=6.400$ $ \mu$m in the time $\tau$. 
Again, this position is kept constant during the time $\tau$.  The cycle is closed with a symmetric backward process, giving a total cycle time of $4\tau=2$ s. Notice that $\tau$ is much larger than the bead relaxation time ($\gamma/\kappa \approx 80$ ${\rm ms}$ ), so we consider  as equilibrium the last 0.4 s of each time interval in which  the  trap position is kept constant. This periodic  protocol is repeated 400 times at each fixed temperature to obtain enough statistics on the measure of the bead positions. The temperature is increased at different steps depending on the distance from the critical temperature. After each temperature step, we wait 1 min to let the system thermalize before acquiring data. This scheme is repeated up to the achievement of the critical temperature.

\begin{figure}[ht]
\centering
\includegraphics[width=.4\textwidth,trim=0 0cm 0 0cm,clip=true]{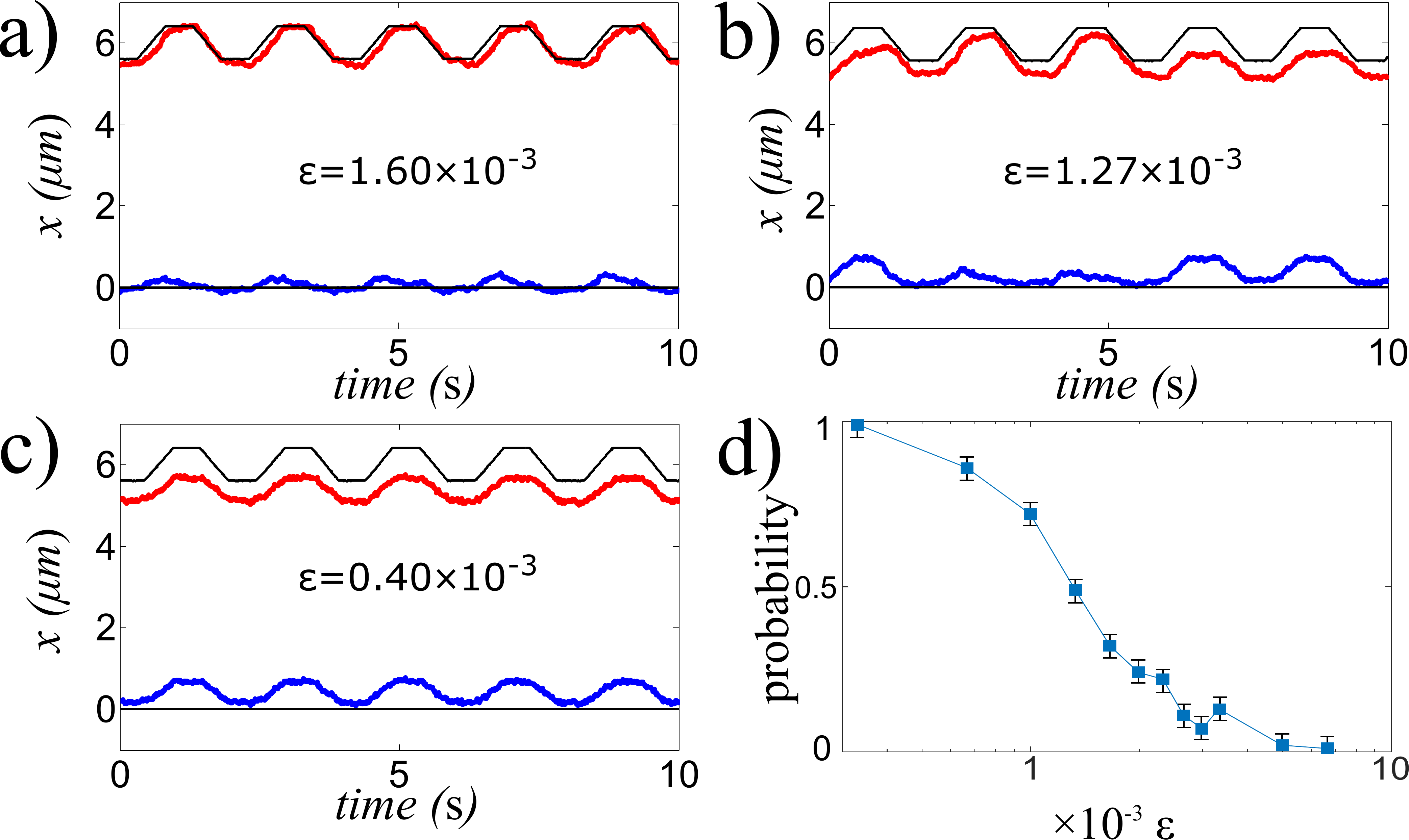} 
\caption{Trajectory of the beads 
 $1$ (blue) and $2$ (red) (\textbf{a}) no synchronization, {$\varepsilon=1.60 \times 10^{-3}$,} 
  (\textbf{b}) weak synchronization, $\varepsilon=1.27 \times 10^{-3}$, (\textbf{c}) complete synchronization, $\varepsilon=0.40 \times 10^{-3}$. The black solid line corresponds to the  position of the optical traps versus time. (\textbf{d}) Probability to be in a synchronous state ($p_{\rm NS}(T)$ ).  Notice how the synchronization increases monotonically when the system approaches the critical temperature. The error is purely statistic $\Delta p_i(T) =1/\sqrt{N(T)}$.
}
\label{fig:fig1}
\end{figure}

In Figure \ref{fig:fig1}, we plot the time evolution of the two beads positions $x_1(t)$ and $x_2(t)$ measured at different $\epsilon$ when the above mentioned protocol is applied. Looking at  Figure \ref{fig:fig1}, we see that particle motion depends on $\epsilon$. Specifically  when $\epsilon$ (Figure \ref{fig:fig1}c) is decreased, the motion tends to synchronize and the mean $x_1$ and $x_2$ are shifted with  respect to the mean  positions of the optical  traps.  Therefore, we use the mean position of the particles during the two equilibrium periods  of the protocols to define if the particles are synchronized (S) or not-synchronized(NS).  More precisely, at each $\epsilon$,  we count the number $\rm{p}_{\rm{S}}$ 
of periods where   the fixed particle is displaced from its normal optical equilibrium position to a new one. We define  the probability $p_{\rm S}(\varepsilon)$ of being synchronized as the  ratio between    
$\rm{p}_{\rm{S}}$ 
and 400, that is the total  number of  times in which the protocol is applied. The probability  $p_{\rm S}(\varepsilon)$ (see Figure \ref{fig:fig1}d) clearly depends on the fluid temperature. When this  gets  close  to $T_C$,  $p_{\rm S}(\varepsilon)$  increases monotonically in this range of temperature, because the critical Casimir force becomes dominant. 

Up to this point we have defined a phenomenological feature: the dynamics of our system depends on the temperature of the surrounding critical mixture. {The next step is to analyze the origin of this  behavior by measuring the  probability density function  $\rho(d)$, where $d=x_2-x_1-2R$ is the distance between the beads surfaces when the trap $2$ is in one of the two equilibrium positions.} From the logarithm of $\rho(d)$ we obtain the total potential $U_{\rm total}(d)$ (Notice that in equilibrium this is valid even  when the dissipation is a function of $d$ as in this case (see Equation (\ref{eq:langRP}) )  .

\begin{figure}[ht]
\centering
\includegraphics[width=.4\textwidth,trim=0 0cm 0 0cm,clip=true]{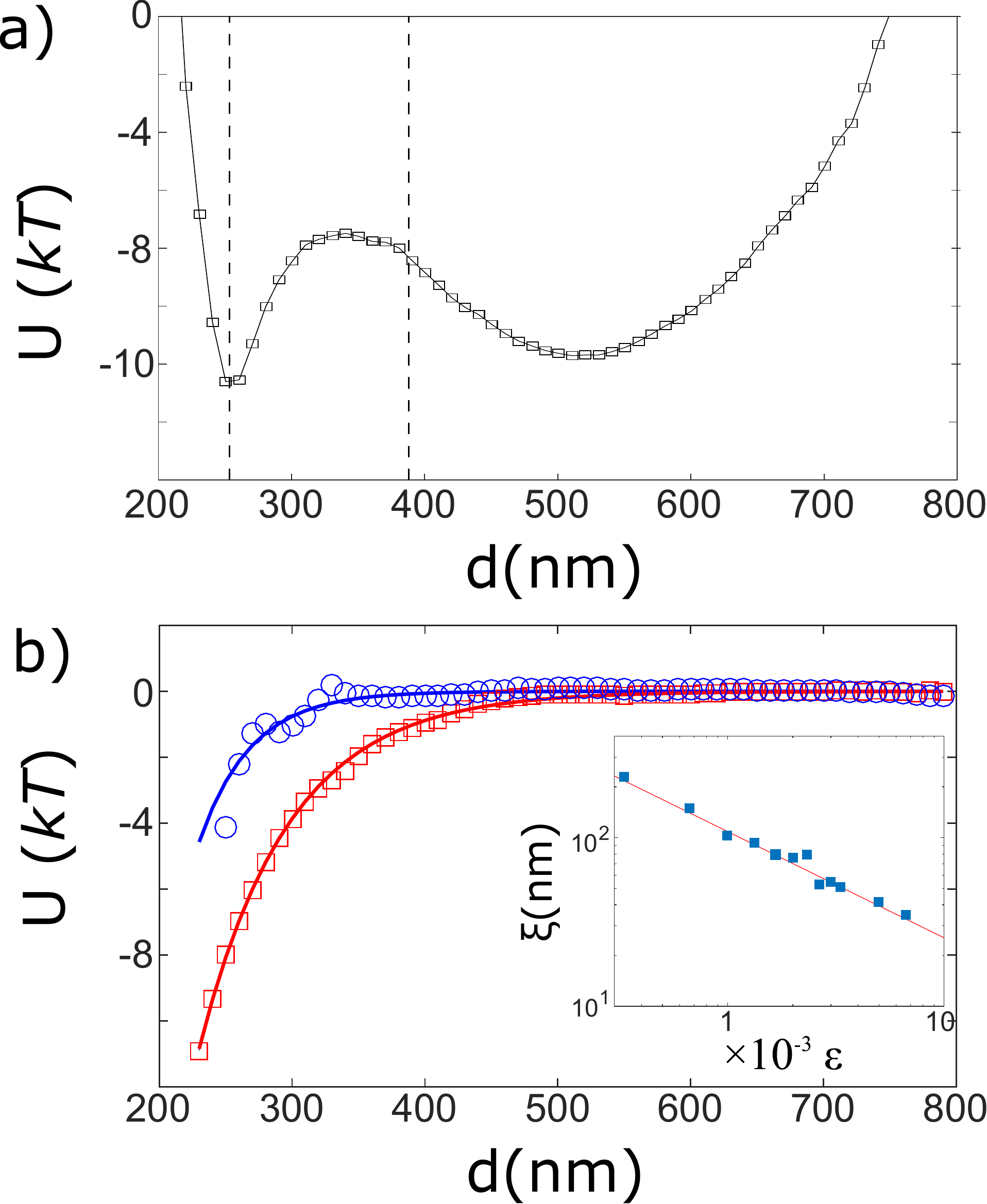} 
\caption{ (\textbf{a}) Total potential $U_{\rm total}(d)$ as a function of the distance between the spheres surfaces $d=x_2-x_1-2R$. The potential has been  measure  using the probability  density function $\rho(d)$ of $d$.  It is possible to divide it in three main parts. At small distances, electrostatic repulsion avoid the sphere to stick together. At further distances the optical potential dominates and creates a local minimum of energy. The Casimir potential changes with temperature and allow the critical force to dominate the dynamic of the system. 
(\textbf{b}) Critical Casimir potential corresponding to $\xi\approx 30$ nm, $T_c-T\approx 2.0$ K, blue circles, and $\xi\approx 70$ nm, $T_c-T\approx 0.5$ K, red squares. Solid lines corresponds to a fit to Derjaguin approximation, Equation (\ref{eq:derjaguin}), keeping the correlation length as a free parameter.  Inset)  The correlation lengths obtained by the previous fit are represented as a function of the reduced temperature (blue solid squares) while the theoretical evolution is represented by the red solid line $\xi=\xi_0 \varepsilon ^{-\nu}$ , where $\varepsilon=(T_c-T)/T_c$ is the reduced temperature, $\xi_0$ $=$ 1.4 nm is the characteristic correlation length of the mixture and  $\nu$ $=$ 0.63 is the universal exponent associated with the transition.}
\label{fig:fig2}
\end{figure}

The total potential can be split   in three main parts: at shortest distances, the electrostatic repulsion between the beads dominates $U_{e}(d)\approx kT\exp{[-(d-l_S)/\sigma]}$  where $\sigma$ is the Debye screening length and $l_S$ depends on the surface charges of the particles. At large  distances the harmonic  potentials of the optical traps  dominate and create a local energy minimum. The optical trapping potential can be assumed as parabolic at these distances, $U_{\rm opt,i}(x_i)=\frac{1}{2}\kappa \left( x_i-x^T_{i}\right)^2$ \cite{martinez2012}. Between them, the Casimir potential defines a local minimum, see Figure \ref{fig:fig2}a). Indeed in our experiment  the Casimir force is attractive because   both  beads  are made of the same material. Thus  their surfaces have the same affinity for the mixture components and this  produces an attractive critical Casimir force  \cite{gambassi2009}.

\section{Discussion}

Under this assumption, we subtract the Casimir interaction from our experimental results, see Figure  \ref{fig:fig2}b. The Casimir potential is evaluated under the Derjaguin approximation for two spheres geometry and symmetric boundary conditions:
\begin{equation}
U_{\rm cas}=-\frac{AR\pi kT}{\xi}\exp\left(-\frac{d}{\xi}\right)
\label{eq:derjaguin}
\end{equation}
where $d=x_2-x_1-2R$ is the distance between the surfaces and $A\approx 1.3$ is a numerical constant from the numerical approximation under Derjaguin approach \cite{gambassi2009}. The only free parameter in this equation is  $\xi$, which is obtained by fitting the experimental data of Figure  \ref{fig:fig2}b). The measured evolution of $\xi$ as a function of temperature is plotted in the inset of  Figure  \ref{fig:fig2}b). The behavior is in agreement with the values  measured from light scattering \cite{corti1984,dietler1988},  from which one estimates  $\xi_0\approx 1.4$~$\rm{nm}$. These results show in a convincing way the contribution of the critical Casimir effect in the observed synchronization. It is important to notice  that  the temperature  range where  we observe the critical effects   is almost one order of magnitude larger than  in previous experiments  \cite{hertlein2008}. This  increase is due to the fact that the  characteristic correlation length of  our mixture  ( $\xi_0\approx 1.4$ $\rm{nm}$)  is much larger than that of water-lutidine, i.e. $\xi_0\approx 0.3$ $\rm{nm}$.

 Let us analyze the energetics of the applied protocols. 
The system has the following forces acting on it: optical, $\vec{F}_{\rm opt}=-\kappa(\vec{x}-\vec{x}^T)$; electrostatic, $\vec{F}_{\rm elec}=kT/\sigma \exp (-(d-l)/\sigma)$; viscous, $\gamma_0 \dot{\vec{x}}$  where $\gamma_0=6\pi\eta R$ and $\eta$  is the dynamic viscosity of the mixture; stochastic, $\vec{\xi}$; Casimir, $\vec{F}_{\rm cas } (r)$, where $r=x_2-x_1$  and $F_{1,\rm cas }=-F_{2,\rm cas }$. It is important to notice that there is a non negligible cross correlation between the particles because the distance between their surfaces, during the experimental protocol, is smaller than the radius of the beads
 \cite{crocker1997}. This cross correlation can be interpreted as a viscosity  gradient, which implies the presence of  a multiplicative noise compensated by an entropic force $F^i_{\rm ent}=T\partial_{x_i} S$, where $S=\frac{k}{2}\log \gamma_r$ and $\gamma_r$ is an $r$ dependent damping (see  Appendix \ref{app:viscosity} and References \cite{mannella2011,sancho1982,McClintock,Rubi}).  The dynamical evolution of the two positions is expressed by the coupled Langevin equations:
\begin{equation}
\gamma_0\dot{\vec{x}}=\tilde{\mathcal H}\vec{F}=\tilde{\mathcal H}\left(\vec{F}_{\rm opt}+\vec{F}_{\rm elec}+\vec{F}_{\rm cas}+\vec{F}_{\rm ent}+\vec{\xi}\right)
\label{eq:langevineq}
\end{equation}
where $\tilde{\mathcal{H}}$ is the hydrodynamic coupling tensor \citep{bartlett2001} with dependence on ${\tilde r}=(x_2-x_1)/R$, see Appendix~\ref{app:viscosity}:
\begin{equation}
\tilde{\mathcal{H}}=\begin{pmatrix}
1-15/4{\tilde r}^4 & 3/2{\tilde r}-1/{\tilde r}^3\\ 
3/2{\tilde r}-1/{\tilde r}^3 & 1-15/4{\tilde r}^4\\
\end{pmatrix}
\label{eq:couplingtensor}
\end{equation}

The stochastic force has zero mean and correlation given by $\langle \xi_i(t)\xi_j(t+\tau)\rangle=2kT\gamma_0\left[\tilde{\mathcal{H}}^{-1}\right]_{ij} \delta(\tau)$.

The strong dependence of the dynamics on the hydrodynamical coupling can be simplified using the eigenvectors of the system: the relative motion $r=x_2-x_1$ and the collective motion $\varphi=x_2+x_1$ of the particles. The evolution of the new coordinates can be expressed by two Langevin equations~as:
\begin{eqnarray}
\gamma_{\varphi} (r)\dot{\varphi}&=&-\kappa[\varphi-x^T_{2}(t)]+\xi_{\varphi},\notag\\
\gamma_r (r) \dot{r}&=&-\kappa[r-x^T_{2}(t)]+2F_{\rm elec}(r)-2F_{\rm cas}(r) - \notag \\ 
& &-\frac{kT}{2}\partial_r \log \gamma_{r}+\xi_r  ,
\label{eq:langRP}
\end{eqnarray}
where  $\xi_{\varphi}=\xi_1+\xi_2$ and $\xi_{r}=\xi_2-\xi_1$, with correlations $\langle \xi_r(t)\xi_r(t+\tau)\rangle=4kT\gamma_r (r)\delta(\tau)$, $\langle \xi_{\varphi}(t)\xi_{\varphi}(t+\tau)\rangle=4kT\gamma_{\varphi}(r)\delta(\tau)$ and $\langle \xi_r(t)\xi_{\varphi}(t+\tau)\rangle=0$. The drag terms are defined including the hydrodynamic coupling as $\gamma_{r}=\gamma_0/(\tilde{\mathcal{H}}_{11}-\tilde{\mathcal{H}}_{12})$ and $\gamma_{\varphi}=\gamma_0/(\tilde{\mathcal{H}}_{11}+\tilde{\mathcal{H}}_{12})$. Notice that both viscosities depend on the relative position $r$, but not on the position of the center of mass because of the isotropy of the surrounding fluid.

\begin{figure}[ht]
\centering
\includegraphics[width=.4\textwidth,trim=0 0cm 0 0cm,clip=true]{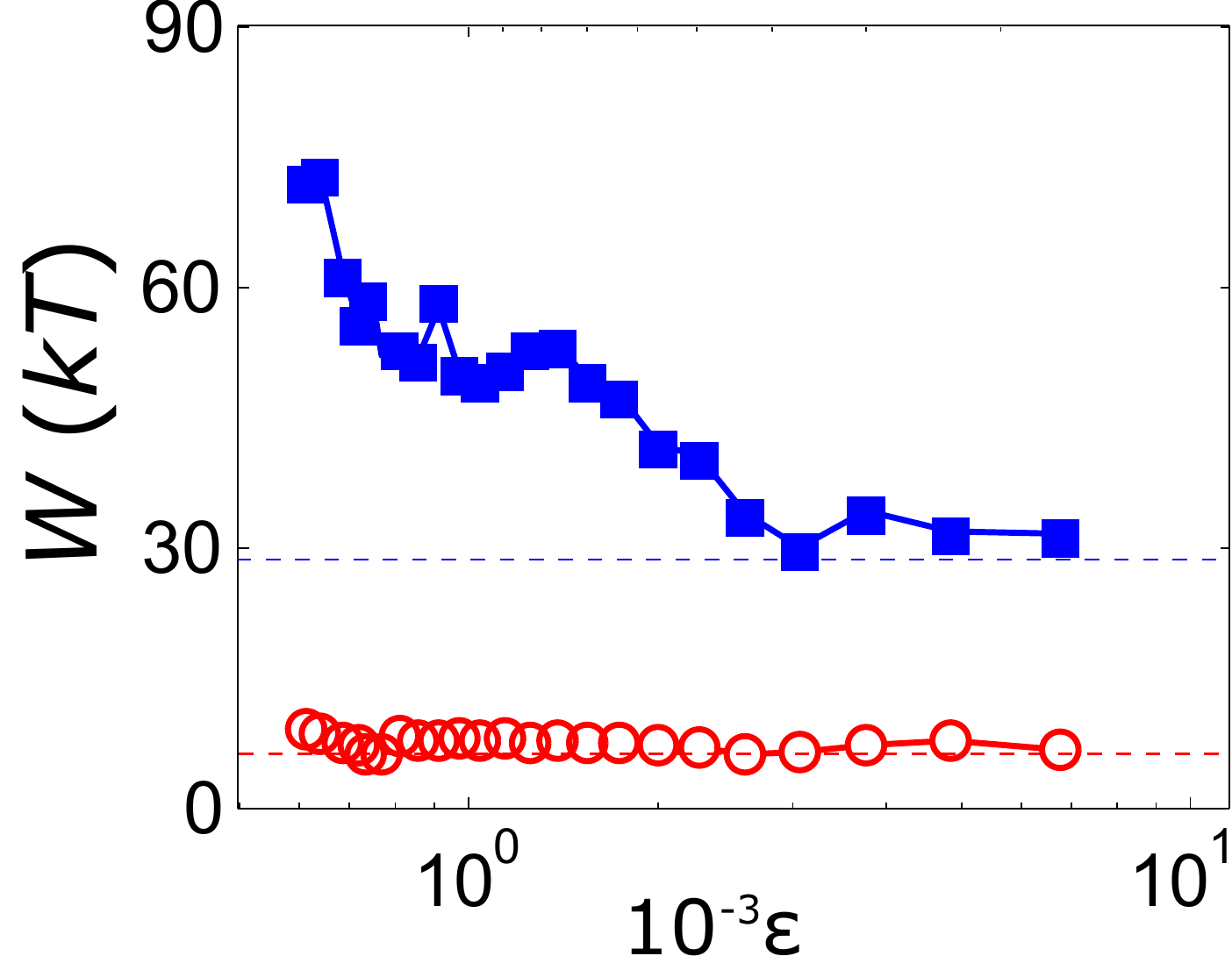} 
\caption{Energetics of the system as a  function of the reduced temperature $\varepsilon$. Work $\langle W_{\varphi}\rangle$ produced in  the collective motion of the particles $\varphi$ (red empty circle). Work $\langle W_r\rangle$ in the relative motion of the particles $r$ (blue solid squares).  $\langle W_{\varphi}\rangle$ shows no dependence with the Casimir force, due to its purely dissipative nature. On the other hand, $\langle W_r\rangle$ has a dependence in $\varepsilon$, as the Casimir force starts to dominate the dynamics. Horizontal dashed lines corresponds to the pure dissipative work in each coordinate. Notice how the higher value of the relative viscosity appears  in a higher value of $\langle W_r\rangle$ even far from the critical point.}
\label{fig:fig3}
\end{figure}

Once the change of coordinates is done, we must point out two important features in the new Langevin equations. First, the Casimir and electrostatic forces only appears explicitly in the relative coordinate, even if the global dynamics is linked with them. Second, the only variable with an explicit  dependence on time is the position of the moving trap $x^T_2$. Within  the stochastic energetics framework \cite{seki2010}, we can define the work exerted on (by) the system as the change of energy produced  by the  external parameters. The differential of the work in each coordinate during the given process is {defined as:} 
\begin{equation}
\delta W_\delta=-\kappa (x_\delta-x^T_{2})\circ\text{d} x^T_{2},
\label{eq:work}
\end{equation}
where the subindex $\delta$ corresponds to each coordinate $\left( r,\varphi\right)$ and $\circ$ denotes the Stratonovich integration \cite{seki2010}. The external force $-\kappa (x_\delta-x^T_{2})$ is directly measured from the displacement of each particle from the equilibrium point. The values of work are averaged on  the 400 trajectories to obtain the average value $\langle W_{\delta}\rangle$ in the both coordinates, see Figure  \ref{fig:fig3}. 

Then, we can combine the definition of work, Equation (\ref{eq:work}), with the Langevin equations, Equation (\ref{eq:langRP}), to obtain a global view of the evolution of the system energetics. We can split the total work in a dissipative part and in a part associated to the external force:
\begin{eqnarray}
W_{\varphi}&=&\int_{\Gamma}(\gamma_{\varphi}\dot{\varphi}-\xi_{\varphi})\circ\text{d}x^T_{2},\\
W_r&=&\int_{\Gamma}2F_{\rm cas}\circ\text{d}x^T_{2} +\\ 
& &+\int_{\Gamma}(\gamma_r\dot{r}+2F_{\rm elec}(r)-\frac{kT}{2}\partial_r \log \gamma_{r}-\xi_r)\circ\text{d} x^T_{2}\notag,
\label{eq:workarray}
\end{eqnarray}
where $\Gamma$ defines the imposed protocol, that, in our particular case, moves the particles from a close to a further position  at a velocity $v=1.70$ $\mu$m/s.  

As $\langle W_{\varphi}\rangle$ has only a dissipative nature, there is no change in the free energy associated to this coordinate during the protocol. Then, we can obtain a value of this energy as: $\langle W_{\varphi}\rangle=\int_{\Gamma}(\langle \gamma_{\varphi}\dot{\varphi}-\xi_{\varphi}\rangle)\circ\text{d}x^T_{2}\approx\gamma_{\varphi}vL=8.6$ {kT}, which is in agreement with the observed value $\langle W_{\varphi}\rangle=(8.0\pm 0.5)$ {kT}
. The case of $\langle W_r\rangle$ is different, because  the  free energy increases when the traps are  taken away and it decreases  when they are approached. This fact is shown in Figure  \ref{fig:fig3}, where the mean values of the work when the beads are taken away is plotted as function of the reduced temperature.  We see how the energetics of the collective motion is not significantly affected by the Casimir force, while the relative work increases when the critical interaction arises. The change in the exerted work agrees with the increase of the depth of the Casimir potential, once we take into account the dissipated work along the protocol in the relative position coordinate.

The fact that the collective motion is not significantly affected by the Casimir force is in agreement with the implicit assumption that the global dynamics of the system is not changing when approaching the criticality. Indeed, the critical exponent for the viscosity is known to be very small, about 0.036 \cite{burstyn1983dynamic}.  We tried to measure the critical change of the viscosity with the beads motion, but the effect is very small and remains within the experimental error.  Thus we can safely conclude that the  critical heterogeneities affect  the interaction through the  Casimir force but not the hydrodynamic coupling. 
Finally one may wonder whether the electrostatic interaction and the stiffness of the trap are affected by the critical point. We carefully checked that  these two quantities remain constant within error bars in the temeperature range of our experiment.
\section{Materials and Methods}

Our experiments are done in a low critical temperature micelle-solvent solution, ${\rm C_{12}\rm E_5}$  in milliQ water at 1.2$\%$ mass concentration.  The sample is always prepared under nitrogen atmosphere to prevent external contamination. This mixture has a correlation length of $\xi_0\approx 1.4$ nm and a critical temperature $T_{C}\approx(30.5 \pm 0.1)$ $^{\rm o}$C \cite{corti1984,dietler1988}. Few microspheres (Fluka silica,  $R=(2.50\pm0.35)$ $\mu$m) per milliliter are added to the final mixture in a low concentration to allow long time measurement without interference. The mixture is injected into a custom made cell $100 \mu$m thick and mechanically sealed to avoid contamination.
Within the fluid cell, the two optical traps are created by a near infrared laser beams (LaserQuantum  $\lambda =$ 1064 nm ) which is focused thanks to a high NA immersion oil objective (Leica $\times$ 63, NA $=$ 1.4). The laser beam position is controlled by an acousto optical  deflector (AA optoelectronics) which  allows us  to  create two different traps  using the time sharing regime at 5 kHz as well as to change their relative positions. One of the two position  is kept fixed (1) and the other is periodically moved (2).  The two optical trap are kept 15 $\mu$m from the cell bottom slide.
The beads images  are acquired   by a high speed camera (Mikrotron MC1310) and their positions are tracked in real time by a suitable software. The tracking resolution is $\pm$5 nm. The acquisition frequency is fixed at 500 frames per second for all experiments. The images of the camera are also used to precisely determined the critical temperature close to the two particles as explained in Appendix~\ref{app:critical}. 

The temperature is controlled by a double feedback system one on the objective and one inside the cell. As the system uses a high NA objective, the cell is in contact with the objective via the immersion oil. Without the second feedback, the objective would act as a thermal bath at lower temperature, creating a temperature gradient within the cell. Temperature is registered with two independent sensors (Pt 1000 $\Omega$) and sent to a programmable temperature controller (Stanford research instruments). The objective and the cell are heated with heater mats (\textit{Minco} 40 $\Omega$ and 80~$\Omega$ respectively). The whole system is isolated from environment  by a box to reduce the effect of  environmental  perturbations both on the position of the particles and on the temperature.

\section{Conclusions}

In conclusion, we have shown how critical Casimir interactions can be used to manage a \textit{toy machine} where colloidal particles play the role of parts of the mechanism. The strong dependence on temperature when we approach  the critical value allows us to use  temperature as a switch that opens or close a mechanical circuit. Thanks to the long intrinsic correlation length of the binary mixture $\rm C_{12}E_5$-water $\xi_0\approx 1.4$ $\rm nm$, we are able to observe a critical Casimir potential further from the critical temperature. In addition, the energetics of the system is studied within the stochastic energetics framework. The mean work changes as a  function of the temperature along one of the system coordinates, probing in an independent way the importance of the critical interaction. We expect that our results can contribute to the raising development of nanotechnologies, allowing the control of  micrometer systems with exquisite accuracy. We can imagine the possibility to implement this interaction in combination with local changes of temperature \cite{martinez2013,berut2014} within a more complex~mechanism.

\acknowledgments{This research has been developed with financial support from the European Research Council Grant OUTEFLUCOP. Authors acknowledge Alberto Imparato and Emmanuel Trizac the solution of the Fokker Planck equation in the presence of a multiplicative noise. Ignacio A. Mart\' inez acknowledges fruitful discussions with Antoine Berut about hydrodynamic interactions in the mesoscale. Ignacio A. Mart\' inez acknowledges  financial support from Spanish Government TerMic (FIS2014-52486-R).}
\bibliography{references_sync}

\clearpage
\section{Appendix}
\subsection{Viscosity dependence in the stationary solution of the Fokker Planck equation.}\label{app:viscosity}
The confinement of the fluctuating field $\phi$ requires to  push the  particles at distances where the hydrodynamics interactions play a key role in the system' s dynamics. The system has the following forces acting on it: optical, $\vec{F}_{\rm opt}=-\kappa(\vec{x}-\vec{x}^T)$, electrostatic, $\vec{F}_{\rm elec}=kT/\sigma \exp (-(d-l)/\sigma)$; viscous, $\gamma_0 \dot{\vec{x}}$  where $\gamma_0=6\pi\eta R$ and $\eta$ is the dynamic viscosity of the sample; stochastic, $\vec{\xi}$; Casimir $\vec{F}_{\rm cas } (x_2-x_1)$, where $F_{1,\rm cas }=-F_{2,\rm cas }$. As the distance between the beads  surfaces is smaller that the radius there is  a non negligible cross correlation between the particles motions. The dynamical evolution of the two positions is expressed by the coupled Langevin equation:
 \begin{equation}
\gamma\dot{\vec{x}}=\tilde{\mathcal H}\vec{F}=\tilde{\mathcal H}\left(\vec{\nabla U} -T\vec{ \nabla S}+\vec{\xi}\right)
\label{eq:langevineq}
\end{equation}
where $U$ is the sum of all the different potentials acting in the system and $S$ is the entropy associated to the viscosity gradient. The stochastic forces has zero mean and correlation given by $\langle \xi_i(t)\xi_j(t+\tau)\rangle=2kT\gamma\left[\tilde{\mathcal{H}}^{-1}\right]_{ij} \delta(\tau)$.
The Langevin equation requires the hydrodynamic coupling tensor  $\tilde{\mathcal{H}}$ to introduce the high correlation between the motions of the particles. The tensor is symmetric and is expressed as follows up the fourth order \citep{bartlett2001}:
\begin{equation}
\tilde{\mathcal{H}}=\begin{pmatrix}
1-15/4{\tilde r}^4 & 3/2{\tilde r}-1/{\tilde r}^3\\ 
3/2{\tilde r}-1/{\tilde r}^3 & 1-15/4{\tilde r}^4\\
\end{pmatrix}
\label{eq:couplingtensor}
\end{equation}

\begin{figure}[!ht]
\centering
\includegraphics[width=.40\textwidth,trim=0 0cm 0 0cm,clip=true]{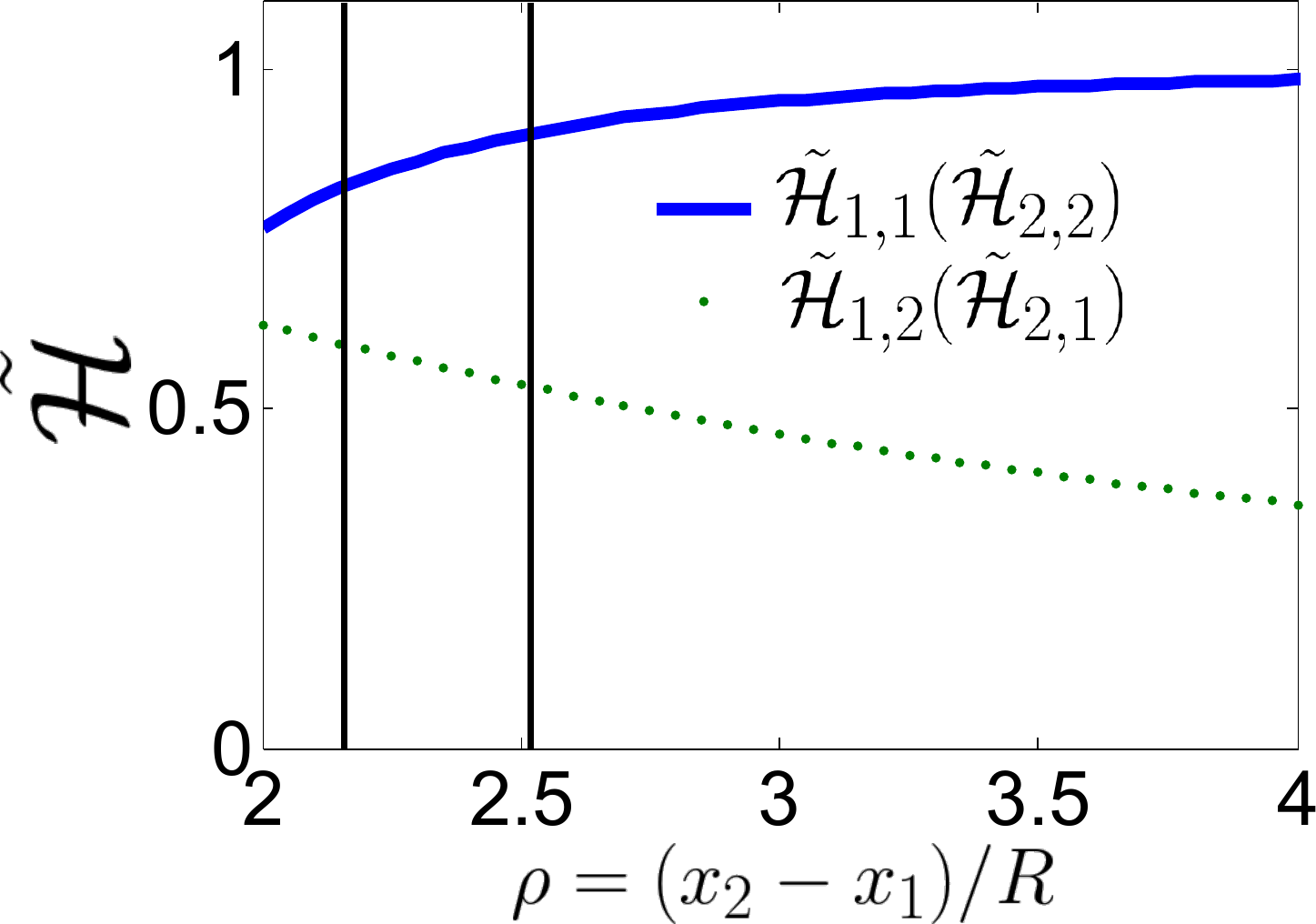} 
\caption{Coupling tensor as a function of the distance. Diagonal (blue solid line) and non diagonal (green dashed line) are represented as a function of the distance between the surfaces over the radius of the bead. The vertical solid black lines represent the interval where our experiment is performed. The hydrodynamic coupling is non negligible in this range.}
\label{fig:couptensSI}
\end{figure}

Figure  \ref{fig:couptensSI} shows the magnitude of both components, diagonal and non diagonal, in the range of our experiment. 

The strong dependence of the dynamics on the hydrodynamical coupling can be simplified using the eigenvectors of the system: the relative motion $r=x_2-x_1$ and the collective motion $\varphi=x_2+x_1$ of the particles. From Equation (\ref{eq:langevineq}), the evolution of the new coordinates can be expressed by two Langevin equations:
\begin{eqnarray}
\gamma_{\varphi} \dot{\varphi}&=&-\kappa[\varphi-x^T_{2}(t)]+\xi_{\varphi},\\
\gamma_r \dot{r}&=&-\kappa[r-x^T_{2}(t)]+2F_{\rm elec}(r)-2F_{\rm cas}(r) - \\ 
& &-T\partial_r S+\xi_r \notag,
\label{eq:langRPap}
\end{eqnarray}
where  $\xi_{\varphi}=\xi_1+\xi_2$ and $\xi_{r}=\xi_2-\xi_1$, with correlations $\langle \xi_r(t)\xi_r(t+\tau)\rangle=4kT\gamma_r\delta(\tau)$, $\langle \xi_{\varphi}(t)\xi_{\varphi}(t+\tau)\rangle=4kT\gamma_{\varphi}\delta(\tau)$ and $\langle \xi_r(t)\xi_{\varphi}(t+\tau)\rangle=0$. Both drag terms are redefined taking into account the hydrodynamic coupling, which makes the two terms different: 
\begin{eqnarray}
\gamma_{r}=\frac{\gamma_0}{\tilde{\mathcal{H}}_{11}(r)-\tilde{\mathcal{H}}_{12}(r)}=\frac{\gamma_0}{1-3/2{\tilde r} +1/{\tilde r}^3-15/4{\tilde r}^4 },\\
\gamma_{\varphi}=\frac{\gamma_0}{\tilde{\mathcal{H}}_{11}(r)+\tilde{\mathcal{H}}_{12}(r)}=\frac{\gamma_0}{1+3/2{\tilde r} -1/{\tilde r}^3-15/4{\tilde r}^4 },
\label{eq:viscDiag}
\end{eqnarray}

\begin{figure}[ht]
\centering
\includegraphics[width=.4\textwidth,trim=0 0cm 0 0cm,clip=true]{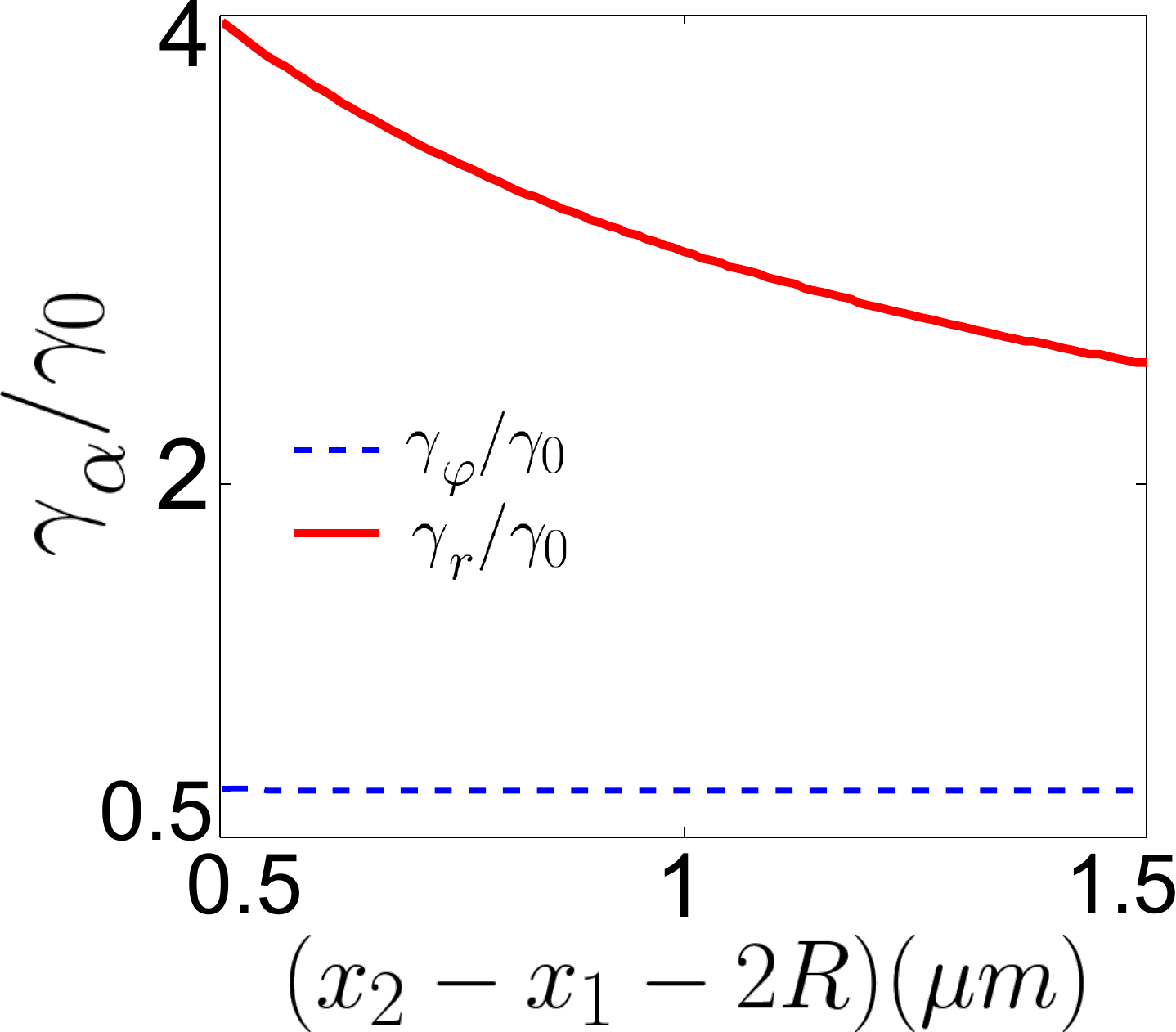} 
\caption{Drag term redefined in the new system of coordinates as a  function of the distance between the surfaces. The collective drag term (blue dashed line) is almost constant in the experimental  range, while the relative drag term (red solid line) has a non negligible variation in the temperature  range. }
\label{fig:diagViscosity}
\end{figure}

In the case of the collective motion $\varphi$, the terms of the expression compensate,  making possible to consider $\gamma_{\varphi}$ as a constant value. On the other hand, the drag term associated to the relative motion has a strong dependence on the relative position, see Figure  \ref{fig:diagViscosity}. 

Next step is to write down the associated Fokker Planck equation. Let us work with the equation associated with the relative position. Being coherent with the selected stochastic integration, Stratonovich integral, and having a spatial dependence of  the viscosity,  $\gamma_r(r)=1/\Omega(r)$ for simplicity, the Fokker Planck equation associated with Equation  (\ref{eq:langRPap}) is written as:
\begin{eqnarray}
\partial_t\rho(r,t) &=& \partial_r \left[\left(\Omega(r)\partial_r (\tilde{U}(r)-TS)\right)\rho(r,t) \right] + \notag \\
 & &-\partial_r\left[\frac{kT\partial_r\Omega(r)}{2} \rho(r,t) \right] + \notag \\
& & +\partial^2_{rr}\left[kT\Omega(r)\rho(r,t)\right],
\end{eqnarray}
where $\rho$ is the position probability density function. For simplicity, the sum of all different conservative forces are grouped in $\tilde{U}(r)$ as $\nabla \tilde{U}\equiv-\kappa[r-x^T_{2}(t)]+2F_{\rm elec}(r)-2F_{\rm cas}(r)$. The stationary solution $\partial_t\rho(r,t)=0$ is:
\begin{equation}
\rho(r)=\frac{\exp\left(-\frac{1}{kT} (\tilde{U}(r)- TS(r))+\log\gamma(r)/2\right)}{\mathcal{Z}}
\end{equation}
which implies that $S=\frac{k}{2}\log{\gamma_r}$ to satisfy Boltzmann distribution.

\section{Determination of the critical point} \label{app:critical}

\begin{figure}[ht]
\centering
\includegraphics[width=.4\textwidth,trim=0 0cm 0 0cm,clip=true]{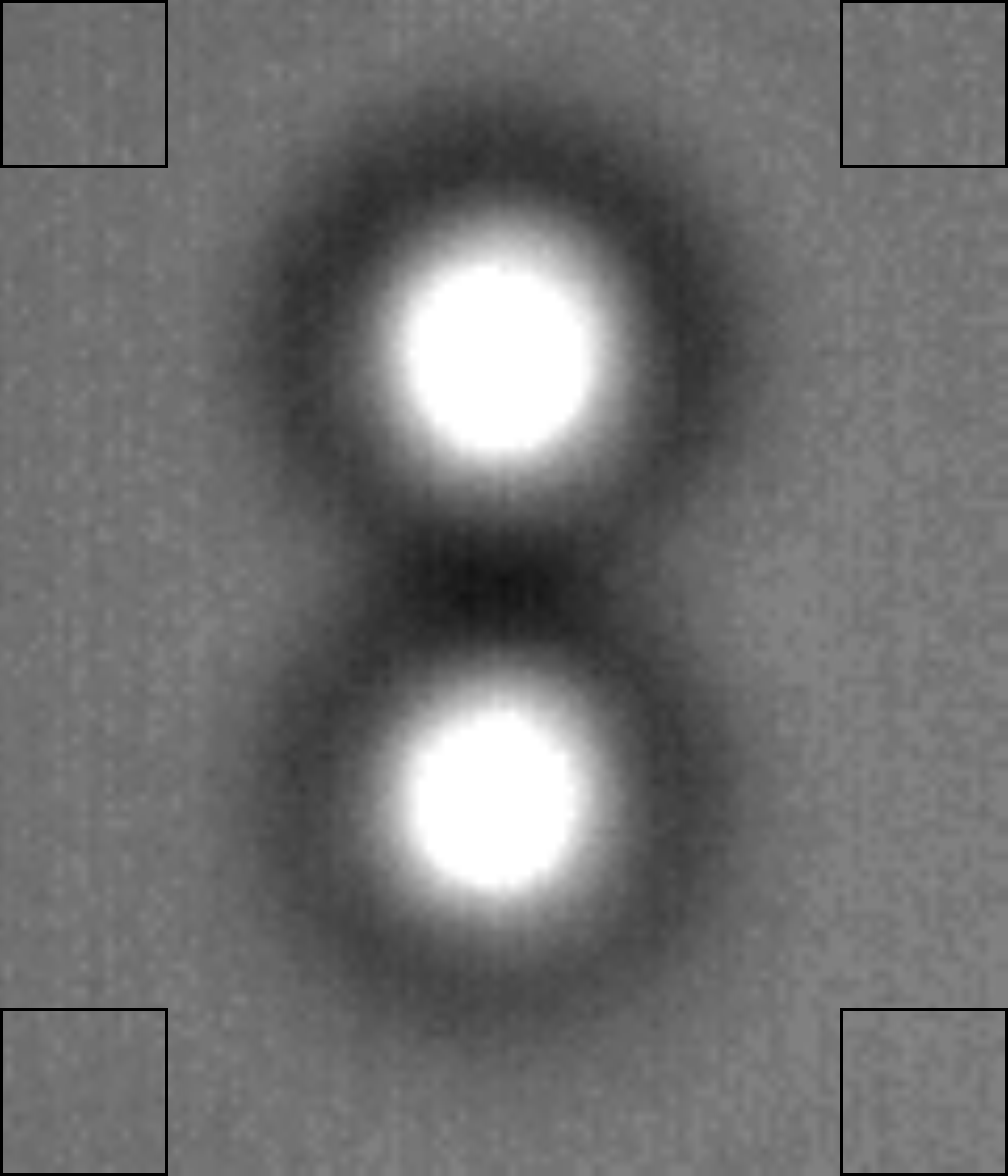} 
\caption{Typical frame of the videocamera used to detect the particle positio In order to characterize the transition point in real time, we analyze the marked square in each of the corners of the image.}
\label{fig:beads}
\end{figure}

One of the major problems in the experimental study of the local properties of the critical binary mixtures close to the critical point, is the interaction with the measurement tools (in our particular case, the optical tweezers) with the sample. In other words, the laser is highly focused in a small spot, making possible a local heating of the sample around the microsystem. On the other hand, the experiment is run in a microfluidic cell, increasing the hysteresis effects associated to the different affinities of the walls of the cell and of the cover-slips. Indeed, once the transition is crossed, the sample does not behave in the same way, needing several hours to have a complete mix of the two components. In order to know exactly when the system transits, we study the  evolution of the images intensity along the experiment, making a local analogy of the classical experiments of critical temperature characterization. We select four regions of 20$\times$20 pixels each (5.85 $\mu m^2$) localized in the four corners of the images, see Figure  \ref{fig:beads}. 

\begin{figure}[ht]
\vspace{11pt}
\centering
\includegraphics[width=.4\textwidth,trim=0 0cm 0 0cm,clip=true]{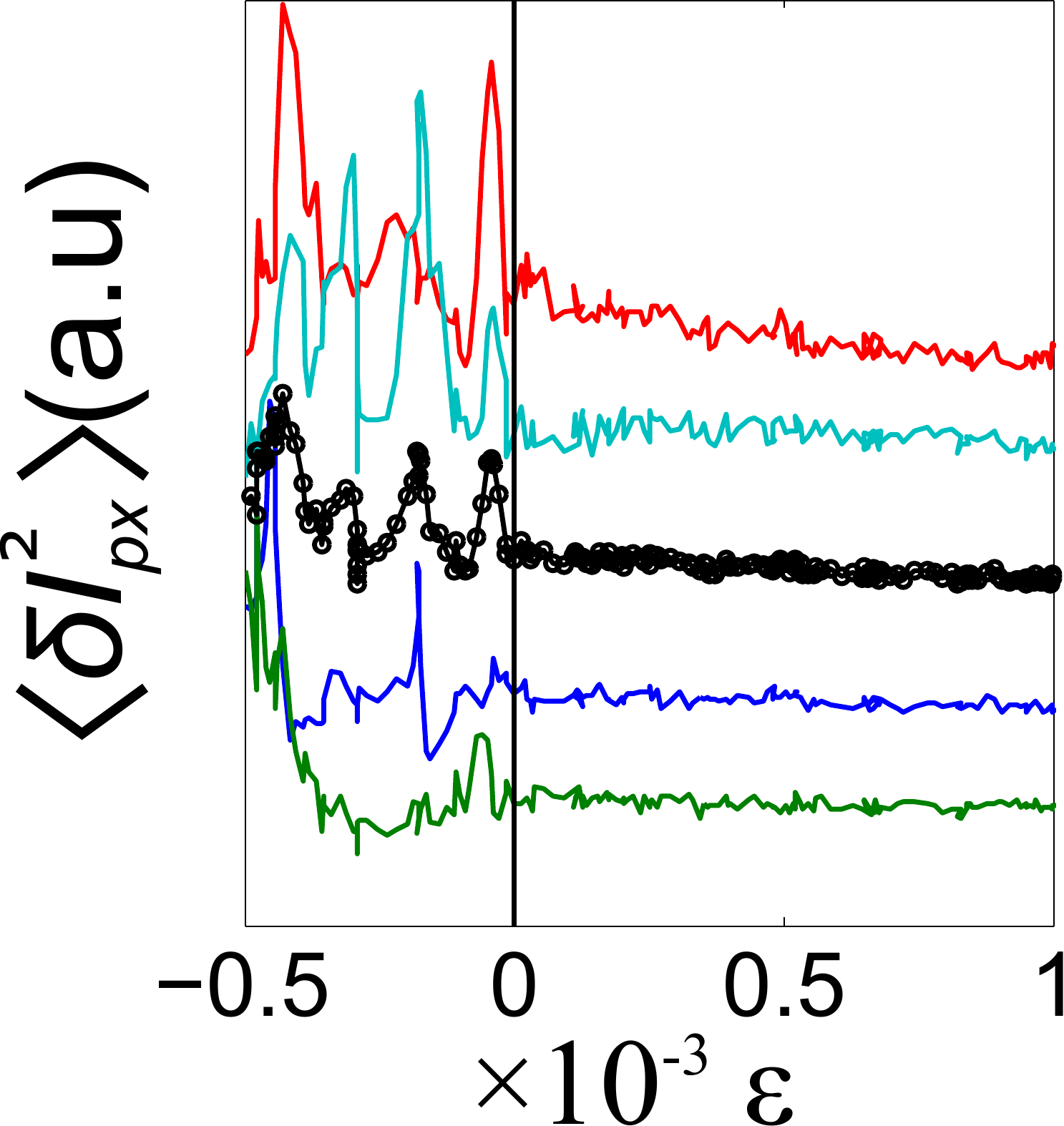} 
\caption{{Variance of the pixels intensities  in each of the regions (solid lines) as a  function of the reduced temperature. The mean value of the four regions is represented with black empty circles. The behavior is almost constant until the liquid goes through  the transition, where the droplets of the different components change the optical properties of the sample.}}
\label{fig:lumSI}
\end{figure}

The variance of the intensity of each region is recorded for each frame and analyzed as a function of the temperature. For an explicit understanding of the technique we show the result that corresponds to an experiment which crosses the critical temperature, see Figure  \ref{fig:lumSI}. { We see that when the critical temperature is crossed, the fluctuations become very large whereas the intensity is almost constant at large $\epsilon$. This effect allows us to identify the transition temperature close to the measurement point}

\end{document}